\documentstyle[preprint,aps,epsfig,tighten,floats]{revtex}

\newcommand{\app}[3]{Astropart.\ Phys.\ {\bf #1}, #3 (#2)}
\newcommand{\hepex}[1]{{\tt hep-ex/#1}}
\newcommand{\hepph}[1]{{\tt hep-ph/#1}}
\newcommand{\astroph}[1]{{\tt astro-ph/#1}}
\newcommand{\prep}[3]{Phys.\ Rep.\ {\bf #1}, #3 (#2)}
\newcommand{\plb}[3]{Phys.\ Lett.\ {\bf B#1}, #3 (#2)}
\newcommand{\npb}[3]{Nucl.\ Phys.\ {\bf B#1}, #3 (#2)}
\newcommand{\cpc}[3]{Comm.\ Phys.\ Comm.\ {\bf #1}, #3 (#2)}
\renewcommand{\apj}[3]{Astrophys.\ J.\ {\bf #1}, #3 (#2)}
\renewcommand{\prl}[3]{Phys.\ Rev.\ Lett. {\bf #1}, #3 (#2)}
\renewcommand{\prd}[3]{Phys.\ Rev.\ {\bf D#1}, #3 (#2)}
\renewcommand{\rmp}[3]{Rev.\ Mod.\ Phys. {\bf #1}, #3 (#2)}
\newcommand{\href}[2]{#1}

\newcommand{\email}[1]{\tt #1}

\begin{document}
	
\begin{titlepage}
\pagestyle{empty}


\title{\hfill {\rm\normalsize MPI-PhT/98-44} \\ \vskip-6pt
  \hfill {\rm\normalsize CfPA-98-Th-12} \\ \vskip-6pt
  \hfill {\rm\normalsize June 1998} \\~\\
  Indirect Detection of Dark Matter in km-size Neutrino Telescopes}


\author{Lars Bergstr\"om}
\address{Department of Physics, Stockholm University, Box 6730,
SE-113~85~Stockholm, Sweden;\\
 E-mail: \email{lbe@physto.se}}

\author{Joakim Edsj\"o}
\address{Center for Particle Astrophysics, University of California,
301 Le Conte Hall, Berkeley, \\
CA 94720-7304, USA; E-mail: \email{edsjo@cfpa.berkeley.edu}}

\author{Paolo Gondolo}
\address{Max Planck Institut f\"ur Physik, F\"ohringer Ring 6, 80805 Munich, 
Germany; \\
E-mail: \email{gondolo@mppmu.mpg.de}}



\maketitle


\begin{abstract}
Neutrino telescopes of kilometer size are currently being planned.  
They will be two or three orders of magnitude bigger than presently 
operating detectors, but they will have a much higher muon energy 
threshold.  We discuss the trade-off between area and energy threshold 
for indirect detection of neutralino dark matter captured in the Sun 
and in the Earth and annihilating into high energy neutrinos.  We also 
study the effect of a higher threshold on the complementarity of 
different searches for supersymmetric dark matter.
\end{abstract}

\end{titlepage}



\section{Introduction} \label{sec:intro}

Neutrino astrophysics will soon enter a new experimental era.  With
the demonstration by the {\sc Amanda} collaboration \cite{amanda} of
the possibility to instrument and successfully deploy kilometer-long
strings with optical modules to 2300 m depth in the ice cap at the
Amundsen-Scott South Pole station, the road to a km$^3$ detector lies
open.  At the same time, endeavors are underway
\cite{nestor,antares,baikal} to prove the possibility of also
deploying a large neutrino detector in the deep ocean (sea), a couple
of years after the termination of the heroic, but not successful,
attempt by the {\sc Dumand} collaboration \cite{dumand}.

Given the fact that such large detectors are likely to be built in the
not too distant future, it appears timely to investigate various
sources of neutrino signals.  Here we will only discuss one very
special source, neutrino-induced high-energy muons pointing back
towards the centre of the Earth or the Sun.  If seen, such a signal
would most likely constitute the indirect detection of weakly
interacting massive dark matter particles (WIMPs), of which
supersymmetric particles, neutralinos are theoretically the most
motivated (for a thorough review, see \cite{jkg} and references
therein).

Supersymmetric neutralinos with masses in the GeV--TeV range are among 
the leading non-baryonic candidates for the dark matter in our 
galactic halo.  One of the most promising methods for the discovery of 
neutralinos in the halo is via observation of energetic neutrinos from 
their annihilation in the Sun \cite{su-an} and/or the Earth 
\cite{ea-an}.  Through elastic scattering with the atomic nuclei in 
the Sun or the Earth, a neutralino from the halo can lose enough 
energy to become gravitationally trapped.  Trapped neutralinos sink to 
the core of the Sun or the Earth where they annihilate into ordinary 
particles: leptons, quarks, gluons and -- depending on their mass -- 
Higgs and gauge bosons.  Because of absorption in the solar or 
terrestrial medium, only neutrinos are capable of escaping to the 
surface.  Neutralinos do not annihilate into neutrinos directly, but 
energetic neutrinos may be produced via hadronization and/or decay of 
the direct annihilation products.  These energetic neutrinos may be 
discovered by terrestrial neutrino detectors.

In a previous work \cite{beg}, we considered neutrino telescopes of 
the presently existing type (such as Baksan, MACRO and 
Super-Kamiokande).  They have a densely instrumented sensitive volume, 
which means that the energy threshold for neutrino-induced muons is 
quite low, on the order of a GeV or even less (a low threshold was 
also assumed in \cite{berez}).  On the other hand their effective area 
is small, no more than about 1000 m$^2$.  The new detectors will have 
much larger area but also higher energy threshold, which motivates a 
second look at this problem.

We have improved our analysis in several important ways since 
\cite{beg}: we have an order of magnitude larger sample 
of allowed supersymmetric models; we have updated the 
experimental and other bounds which define which models are allowed;  
in the calculation of relic density of neutralinos we have incorporated the 
effects of coannihilations as recently computed in 
\cite{coann,jephd}.
 
\section{Muon energy threshold of large detectors}

The new generation of much larger neutrino telescopes, utilizing large
volumes of natural water (ice), will have a much more sparse
instrumentation, with of the order of 100 meters between the
individual strings of optical modules and some 20 m spacing between
the modules on a string.  To get a useful trigger for upgoing events,
needed to suppress the background from downward-going muons generated
by cosmic rays in the atmosphere, several modules on several strings
need to register Cherenkov photons within some specified time window.
From the relative timing of the signals in the modules hit, the
trajectory of the muon can be reconstructed within a few degrees
(depending, among other things, on the energy).

Since a minimum-ionizing muon loses on the order of 2.6 MeV per
centimeter traveled (in water or ice), a horizontal track between two
strings can only be generated by a muon of energy 25 GeV or higher.
Of course, the reconstruction efficiency as a function of energy has
to be computed for a specific detector through a detailed Monte-Carlo
simulation.  Results based on a very preliminary analysis of {\sc
Amanda} 4-string data \cite{bouchta} indicate, however, that below a
certain energy the efficiency drops very rapidly and can for our
purposes be replaced by a step function in muon energy.  In this
paper, we will consider thresholds of 1, 10, 25, 50 and 100
GeV\@. Hopefully, the new detectors will be designed such that the
threshold does not go too much above 25 GeV, but for completeness we
also treat higher values.  As we will show, for the dark matter
detection capability a low threshold is an important design criterion
to be kept in mind when planning the new neutrino telescopes.

If the neutralino mass is above the threshold energy by a fair amount, 
not too many events are lost by increasing the energy threshold.  This 
is because the detection rate is determined by the second moment of 
the neutrino energy (one power of energy because of the rise of the 
neutrino cross section, and one power because of the increasing muon 
path length). So the most energetic muons dominate the rate. A higher
threshold may even be more advantageous than this, because it
also reduces the background from atmospheric neutrinos.  
Thus, a detailed analysis needs a full simulation of both the signal 
and the background, including the angular acceptance window and 
angular resolution, as done in Ref.~\cite{bek} for a 
low threshold.  We will, however, mainly be concerned with the effects 
on the signal of an increased threshold, postponing a more detailed 
analysis to the future, when more details are known about actual 
detectors.

It should be noted that to some extent the threshold energy can be 
lowered at the expense of decreasing the effective reconstructed area 
of the type of neutrino detectors presently being discussed.  For 
instance, when searching for vertically upward-going tracks from the 
centre of the Earth, it may be enough to demand 3 consecutive close 
hits (as defined by signal strength) on a single string.  This selects 
tracks that move almost vertically near one string with an energy 
threshold of not much more than 10 GeV\@. However, this obviously brings 
down the effective area by a substantial amount.  We hope that our 
analysis will be useful when estimating the trade-off between energy 
threshold and signal strength in situations like this, by combining 
our results for rates with detector-specific effective reconstructed 
areas for various thresholds.

We will focus on the fluxes of neutrino-induced muons in units of
km$^{-2}$ yr$^{-1}$. The conversion to an event rate by multiplying by
an effective area is only correct for a thin detector. With detectors
of a volume $\cal O$(km$^3$), the thin approximation is only valid for
muon energies above a few hundred GeV\@. Below that, the contained event
rate will be higher than the event rate of impinging muons. To take
this into account in detail requires a full detector simulation, which
is outside the scope of this paper. We will thus focus on the muon flux
impinging on the detector and show an example of how the event rate
changes for contained events.


\begin{table}
  \begin{tabular}{rrrrrrrr} 
  Parameter & $\mu$ & $M_{2}$ &
  $\tan \beta$ & $m_{A}$ & $m_{0}$ & $A_{b}/m_{0}$ & $A_{t}/m_{0}$ \\
  Unit & GeV & GeV & 1 & GeV & GeV & 1 & 1 \\ \hline 
  Min & -50000 & -50000 & 1.0  & 0     & 100   & -3 & -3 \\
  Max & 50000  & 50000  & 60.0 & 10000 & 30000 &  3 &  3 \\  
  \end{tabular} 
\caption{The ranges of parameter values used in our scans of the MSSM
  parameter space. Note that several special scans aimed at
  interesting regions of the parameter space has been performed.
  In total we have generated about 82000 models that are not excluded by
  accelerator searches.}  \label{tab:scans}
\end{table}

\section{Set of supersymmetric models}

We work in the same minimal supersymmetric standard model with seven 
parameters used in Refs.~\cite{bg,beg,coann,bub,beu}.  In particular, 
we keep a very general set of models imposing no restrictions from 
supergravity other than gaugino mass unification.  For our definition 
of parameters and a full set of Feynman rules, see \cite{bg,jephd}.

We make extensive scans of the model parameter space, some general and 
some specialized to interesting regions.  We vary the model parameters 
within the generous ranges listed in Table~\ref{tab:scans}.

For each model allowed by current accelerator constraints (mainly LEP 
bounds \cite{lepbounds} and constraints from $b\to s\gamma$ 
\cite{cleo}) we calculate the relic density of neutralinos 
$\Omega_\chi h^2$, where $\Omega_\chi$ is the density in units of the 
critical density and the present Hubble parameter is $100h$ km 
s$^{-1}$ Mpc$^{-1}$.  We use the formalism of 
Ref.~\cite{GondoloGelmini} for resonant annihilations and threshold 
effects, keeping finite widths of unstable particles, including all 
two-body tree-level annihilation channels of neutralinos.  A major new 
improvement compared to \cite{beg} is that coannihilations are 
included in the relic density calculations according to the analysis 
of Edsj\"o and Gondolo \cite{coann}.

Present observations favor $h=0.6\pm 0.1$, and a total matter density 
$\Omega_{M}=0.3\pm 0.1$, of which baryons may contribute 0.02 to 0.08 
\cite{cosmparams}.  Not to be overly restrictive, we accept 
$\Omega_\chi h^2$ in the range $0.025$ to $0.5$.  The lower bound is 
somewhat arbitrary as there may be several different components of 
non-baryonic dark matter, but we require that neutralinos are at least 
as abundant as to make up most of the dark halos of galaxies.

\begin{figure}[!t]
\centerline{\epsfig{file=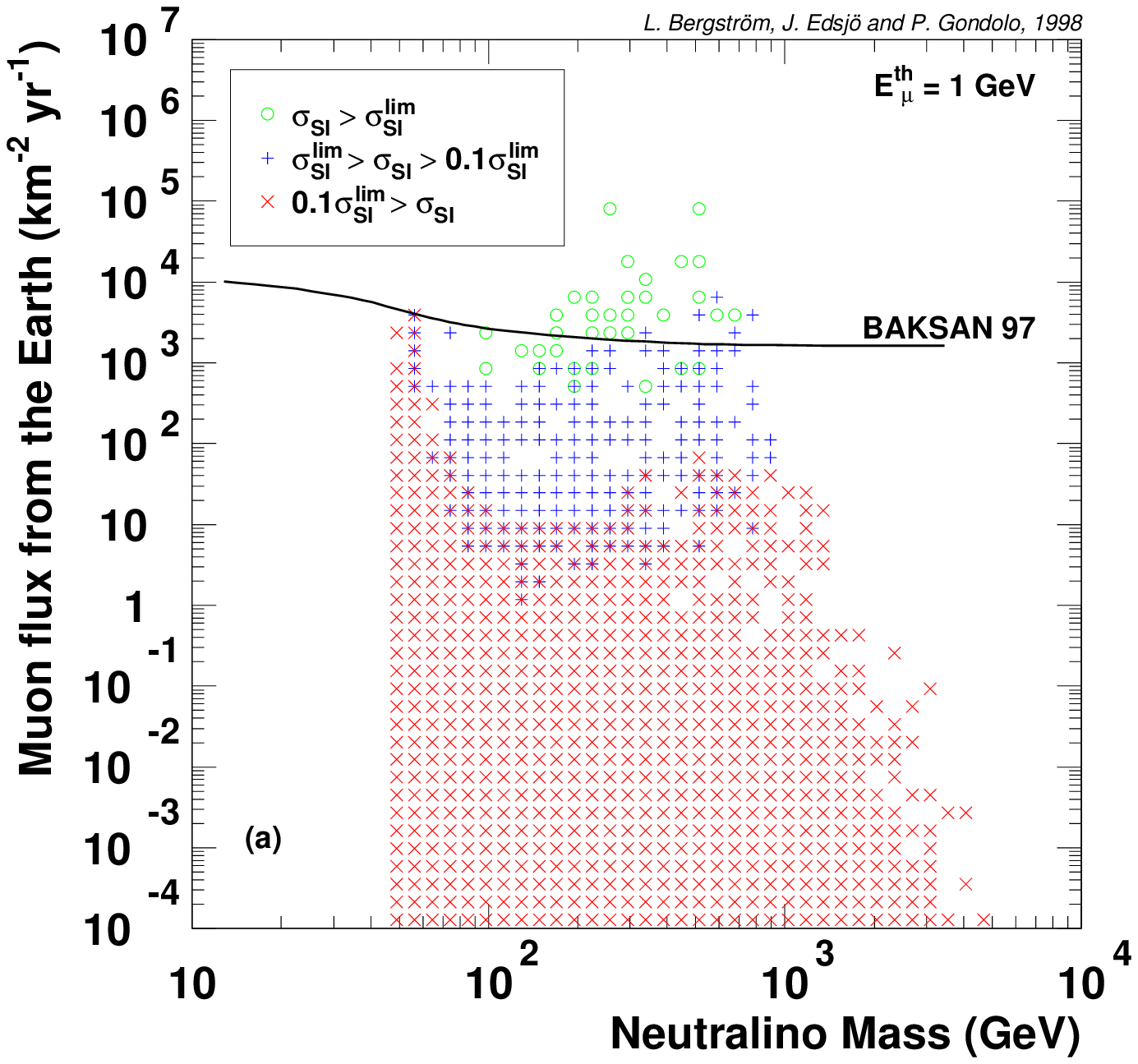,width=0.49\textwidth}
\epsfig{file=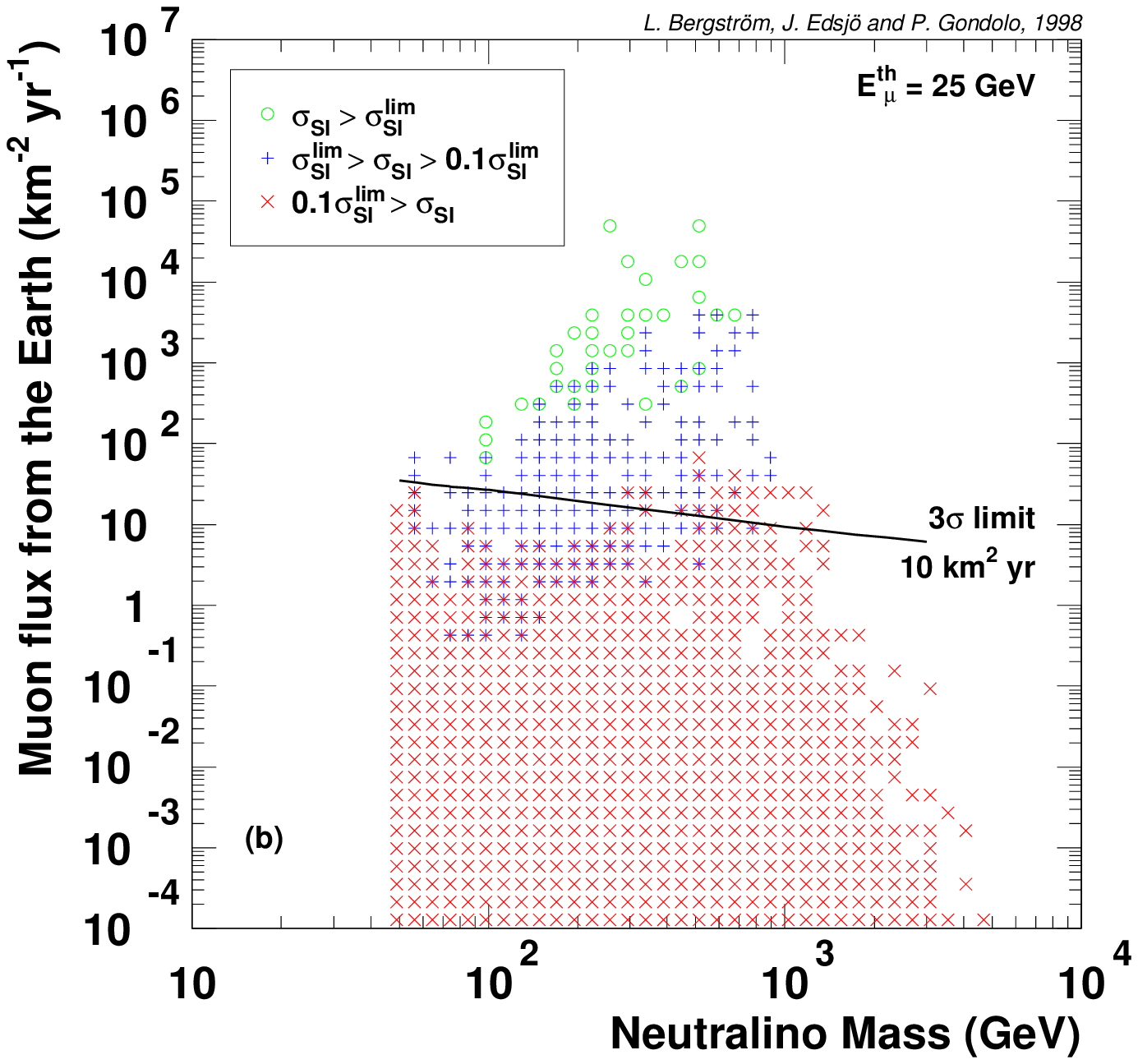,width=0.49\textwidth}}
\caption{The muon fluxes versus the neutralino mass for annihilation
in the Earth. The muon energy threshold is (a) 1 GeV and (b) 25 GeV\@.
The horizontal line is in (a) the current limit from Baksan
\protect\cite{baksan} and in (b) the best limit that can be reached
with a neutrino telescope with an exposure of 10 km$^2$ yr.  Models
already excluded by direct detection experiments and models that will
be explored with a factor of 10 increase in sensitivity are shown with
different symbols (and different colours).}
\label{fig:reamx}
\label{fig:firstfig}
\end{figure}

\begin{figure}[!t]
\centerline{\epsfig{file=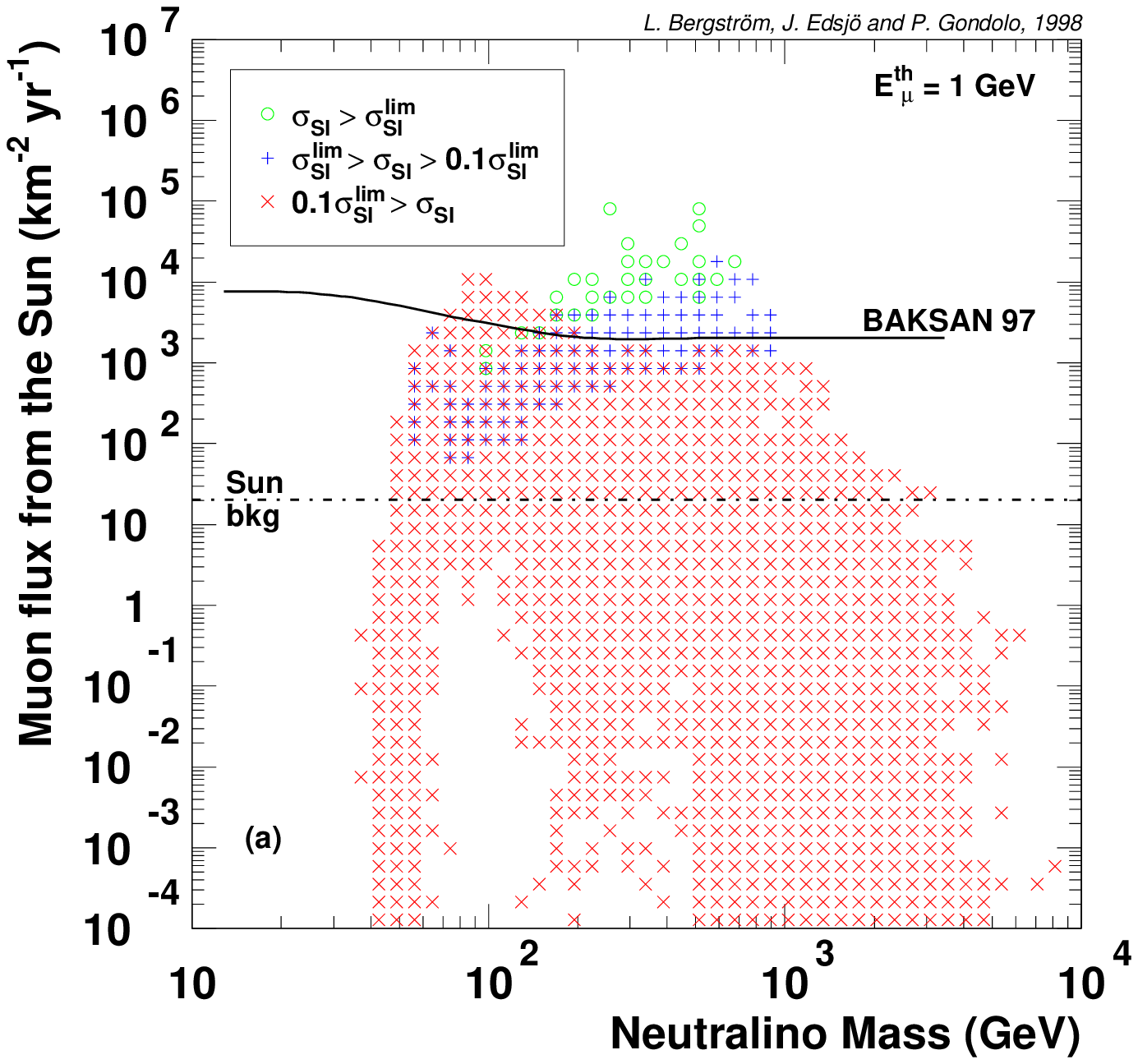,width=0.49\textwidth}
\epsfig{file=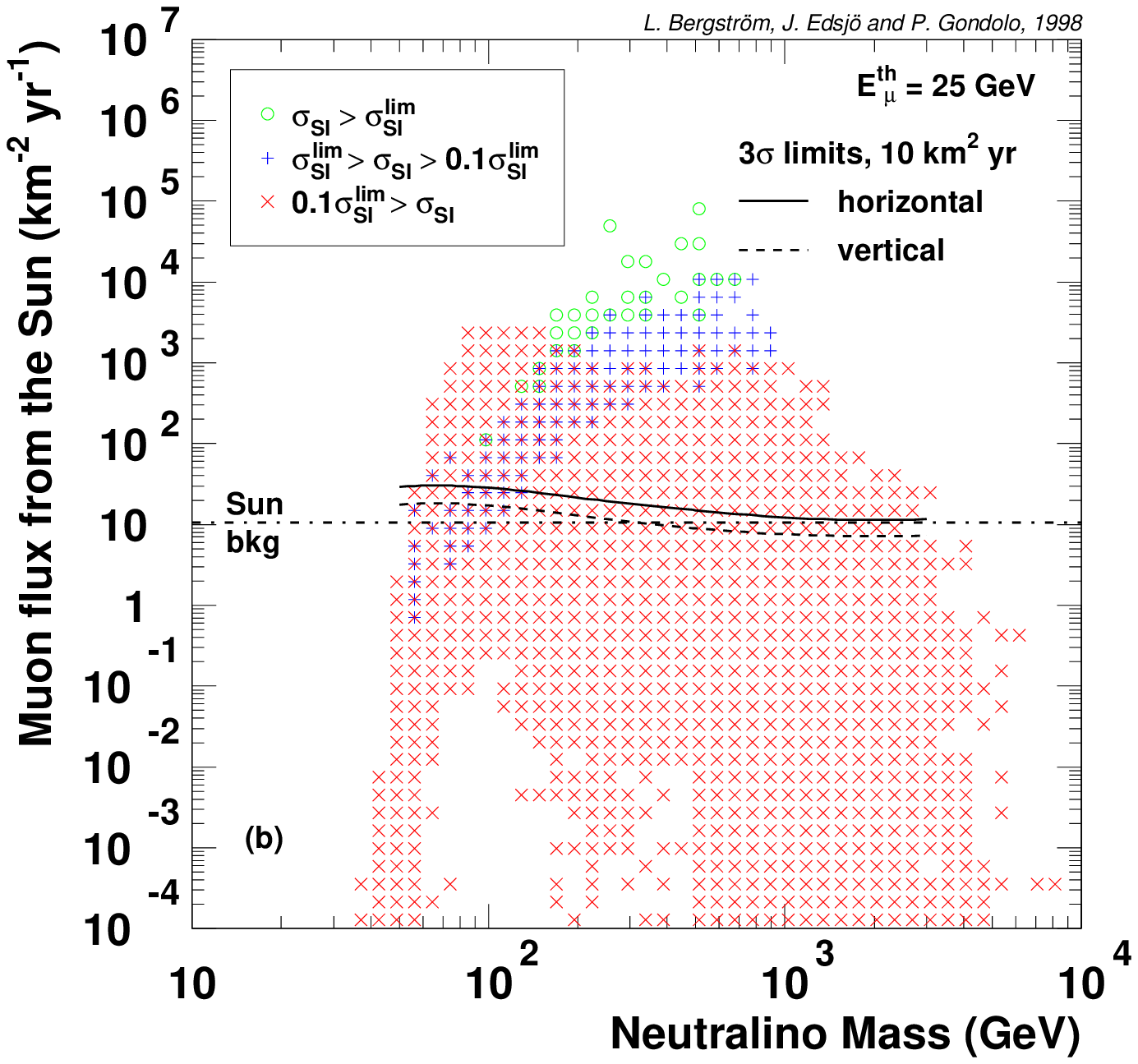,width=0.49\textwidth}}
\caption{The same as Fig.~\ref{fig:reamx}, but for neutralino
annihilation in the Sun. In (b) the best limits with an exposure of 10 
km$^{2}$ yr are given for horizontal and vertical background 
respectively. The expected background from cosmic ray interactions in 
the Sun is also shown.}
\label{fig:rsumx}
\end{figure}

For each allowed model, we compute the different direct and indirect 
detection rates.  Figures \ref{fig:firstfig} to \ref{fig:lastfig} show 
our results.

\section{Muon fluxes from neutralino annihilations}

The capture rate in the Earth is dominated by scalar interactions, and 
presents kinematic enhancements whenever the mass of the neutralino 
almost matches one of the heavy elements in the Earth.  For capture in 
the Sun, both axial interactions with hydrogen and scalar interactions 
with heavier elements are important.  For both the Sun and the Earth 
we use the convenient approximations available in \cite{jkg}.

The prediction of muon rates is quite involved: we compute neutralino 
capture rates in the Sun and the Earth, branching ratios for different 
annihilation channels, fragmentation functions in basic annihilation 
processes, interactions of the annihilation products with the 
surrounding medium (where appropriate), propagation through the solar 
or terrestrial medium, charged current cross sections and muon 
propagation in the rock, ice or water surrounding the detector.  This 
has been performed in the same way as in Ref.\ \cite{beg}.  Recently, Chen 
and Kamionkowski \cite{chenkam} calculated the annihilation cross 
section including three-body final states just below two-body final 
state thresholds.  They concluded that in these specific regions of 
the parameter space, the fluxes can increase by a factor of a few when 
three-body final states are included.  We have not included these 
three-body final states in our calculation since they do not change 
the overall picture presented here.

We simulate the hadronization and/or decay of the annihilation
products, the neutrino interactions on the way out of the Sun and the
neutrino interactions close to or in a detector with {\sc Pythia}
6.115 \cite{pythia}.  We treat the interactions of the heavy hadrons
in the centre of the Sun and the Earth in an approximate manner as
given in Ref.\ \cite{jephd}.  We simulate $1.25 \times 10^6$ events
for each annihilation channel and for each of a set of 18 different
neutralino masses.  We then interpolate in these tables and let Higgs
bosons decay in flight to obtain the neutrino-induced muon flux for
any given MSSM model.

\subsection{Backgrounds and signal extraction}

The most severe background when looking for neutrinos from neutralino 
annihilation in the Sun/Earth is the atmospheric background produced 
by cosmic rays hitting the Earth's atmosphere \cite{atm-nu}.  For the 
Sun, there is also a background from cosmic ray interactions in the Sun 
\cite{sun-bkg} which is small but irreducible (at least as long as energy is 
not measured).

Even though the energy and angular dependence of the atmospheric
background and of the signal are very different, we can essentially
only make use of the angular information with present designs of
neutrino telescopes, which have very poor energy resolution.  One way
to search for an excess of neutrinos from the Sun/Earth is to compare
the measured flux in an angular cone with half-aperture angle $\theta$
towards the centre of the Sun/Earth with the expected background from
atmospheric neutrinos in this cone.  The angle $\theta$ can be
determined as optimally as possible to maximize the signal to
background ratio.  We can, however, use the knowledge we have of the
actual shape of the signal, which can be parameterized as
\begin{equation} \label{eq:param}
     \frac{d^2 \phi_s}{dE d\theta} (E,\theta) = \phi_s^0
     \left[ a f_{\rm hard}(m_\chi,E,\theta)+(1-a) 
     f_{\rm soft}(m_\chi,E,\theta) \right] ,
\end{equation}
where $a$ is a model-dependent parameter describing the `hardness' of
the neutrino-induced muon spectrum, $f_{\rm hard}$ and $f_{\rm soft}$
are generic hard and soft muon spectra and $\phi_{s}^{0}$ is the
normalization of the flux.  A typical hard spectrum is given by the
annihilation channel $W^{+}W^{-}$ and a typical soft spectrum is given
by the annihilation channel $b \bar{b}$.  Integrating above a given
detector's threshold, we only have three unknown parameters
$\phi_{s}^{0}$, $a$ and $m_{\chi}$.  We can either fit for these three
simultaneously or assume that $a$ and $m_{\chi}$ are known if we have
a given set of MSSM models we want to test for.  With this more
detailed analysis, we can put limits that are up to a factor of two
better than just using an optimal cut $\theta_{\rm max}$.  This is
described in more detail in Ref.~\cite{bek} and we will compare our
predicted fluxes with the expected discovery potential for a neutrino
telescope with an exposure of 10 km$^{2}$ yr (or 10 km$^3$ yr for
contained events).  We will for this purpose use the $3\sigma$-limits
that can be obtained under the assumption that only $\phi_{s}^{0}$ is
unknown and that the energy spectrum is hard. If we loosen these
assumptions the limits will be up to a factor of 2--3 higher.

For very high exposures (${\cal E} > 10$ km$^{2}$ yr) towards the Sun,
the above limits will be too optimistic due to the background from
cosmic ray interactions in the Sun's corona. This background will have
about the same angular distribution as the neutralino signal from the
Sun, but quite different energy distribution.  With a neutrino
telescope without energy resolution, this background will put a lower
limit on how small fluxes we can probe from the Sun.  The background
fluxes are about 20, 13, 11, 8.6 and 6.6 muons km$^{-2}$ yr$^{-1}$ for
muon energy thresholds of 1, 10, 25, 50 and 100 GeV respectively. For
contained events the corresponding numbers are 724, 75, 31, 16 and
8.0 muons km$^{-3}$ yr$^{-1}$.  We will show these background fluxes in
the figures for the Sun.

\subsection{Dependence on energy threshold}

We start by giving the results for a threshold of 1 GeV, to make
contact with our previous analysis \cite{beg}.  (Of course, these
results are highly relevant for present and planned neutrino detectors
with low energy threshold.)  Fig.~\ref{fig:reamx} (a) shows the muon flux
above 1 GeV from neutralino annihilation in the Earth as a
function of neutralino mass, and Fig.~\ref{fig:rsumx} (a) the
corresponding result from the Sun.  The present bounds from
the Baksan detector \cite{baksan} are indicated by the almost horizontal
lines. Some models giving the highest rates
are excluded under our assumptions, but  we prefer to keep them in
the plots, since the bounds are not completely watertight.
For instance, they depend on the local halo density which may be
uncertain by at least a factor of 2 (depending, among other things, on
the degree of flattening of the Milky Way halo \cite{guyk}).

Many of the main features displayed in Figs~\ref{fig:reamx} (a) and
\ref{fig:rsumx} (a), valid for a muon detection threshold of 1 GeV,
remain essentially unchanged when increasing the muon detection
threshold. As an example of a probably realistic threshold of a
km-size neutrino telescope, we choose 25 GeV\@. In Figs.~\ref{fig:reamx}
(b) and \ref{fig:rsumx} (b) we show the muon fluxes versus the
neutralino mass for a threshold of 25 GeV\@. We also show the best
limits obtainable with an exposure of 10 km$^{2}$ yr, and for the Sun,
the background from cosmic ray interactions in the Sun's corona. Note
that for high masses, there is no need to go above an exposure of
about 10 km$^2$ yr towards the Sun (unless the detector has good
energy resolution) due to the irreducible background from the Sun's
corona. For lower masses, the corresponding exposure would be $10^2$
km$^2$ yr.

\begin{figure}[!t]
\centerline{\epsfig{file=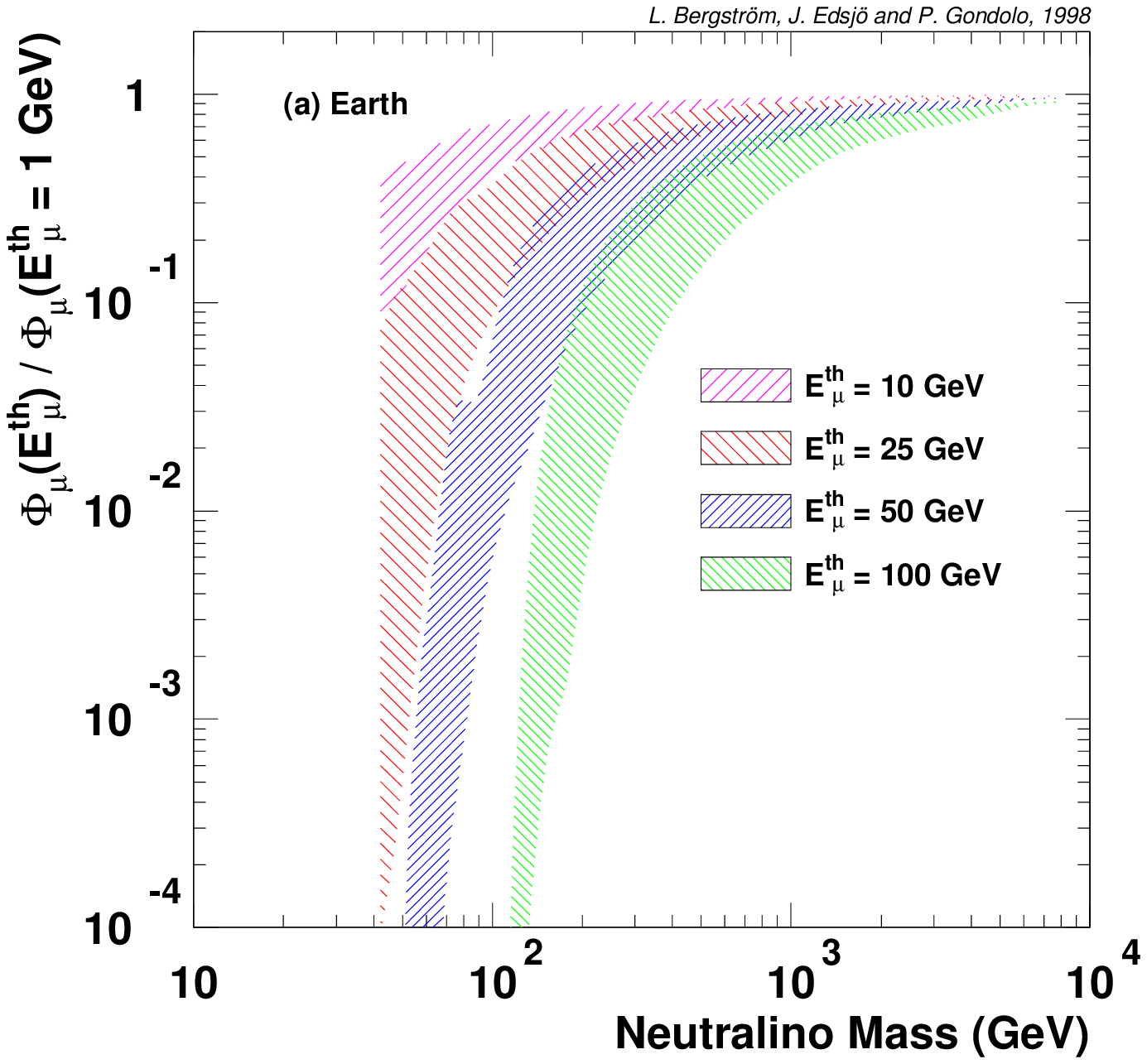,width=0.49\textwidth}
\epsfig{file=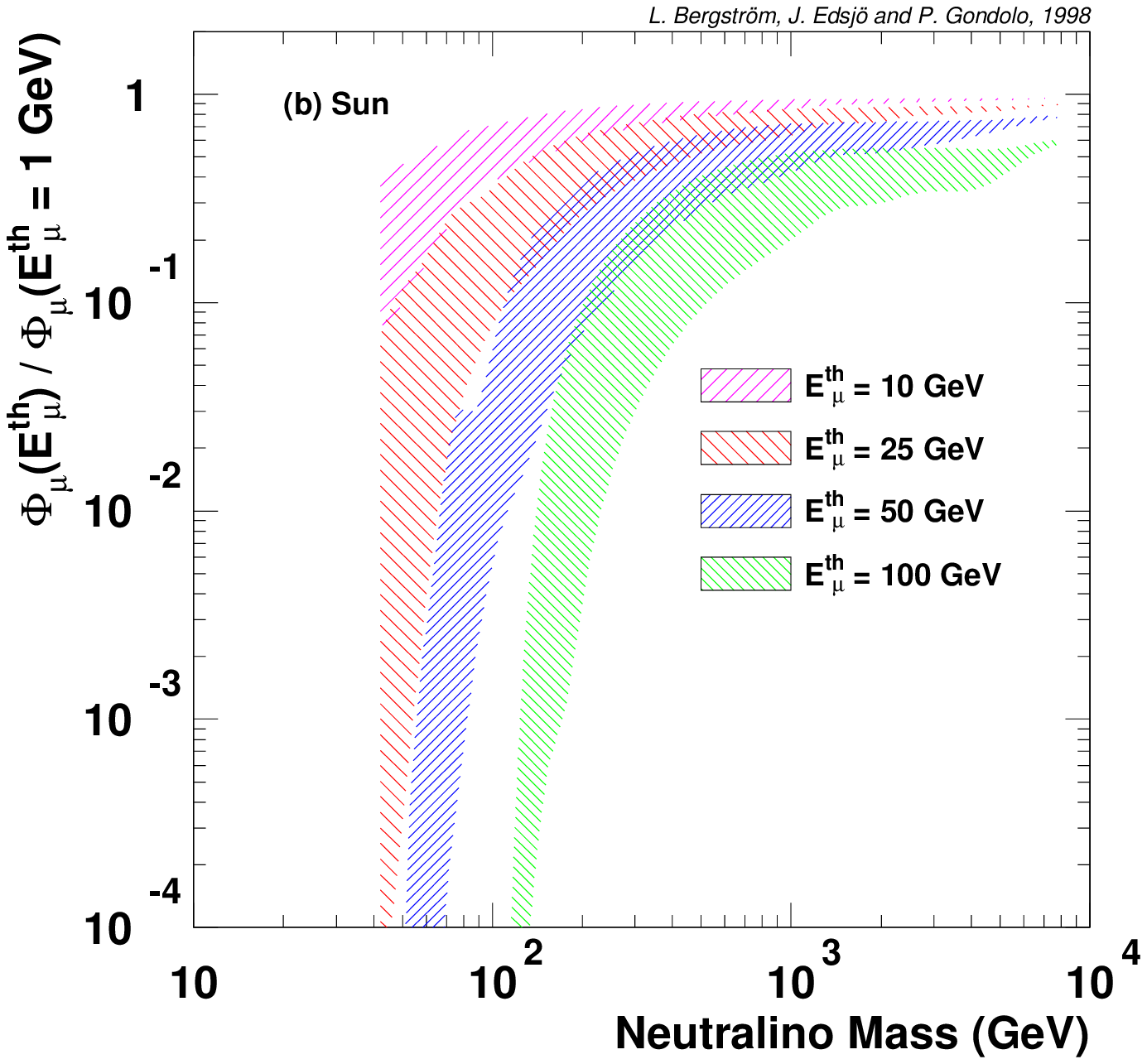,width=0.49\textwidth}}
\caption{The ratio of the muon fluxes with a threshold of $E_\mu^{\rm th}$ 
to those with a threshold of 1 GeV in (a) the Earth and (b) the Sun versus
the neutralino mass. For each given threshold, a band of allowed 
ratios is given.}
\label{fig:reasurmx}
\end{figure}

In Fig.~\ref{fig:reasurmx} we show the ratio of the muon flux with 
different thresholds $E_{\mu}^{th}$ to those with a threshold of 1 
GeV\@. The width of the bands reflects the different degrees of softness 
of the neutrino spectra for a given neutralino mass.  The softer the 
neutrino spectrum, the more we lose by increasing the threshold.  In 
the case of the Sun, we see that there is always a loss even at the 
highest masses.  This is due to the absorption of neutrinos in the 
interior of the Sun, which softens the neutrino spectra.  When the 
threshold exceeds 100 GeV, at least half of the signal from the Sun is 
lost whatever the neutralino mass.

As an example, if the neutralino mass were 200 GeV, increasing the 
threshold from 1 to 100 GeV could decrease the signal by a factor of 
between 10 and 1000.  Fig.~\ref{fig:reasurmx} includes spectra from 
all MSSM models we generated (i.e.\ is not a soft or hard 
approximation).  On the other hand, if the threshold can be kept at 25 
GeV or below, we see that on the average only a factor of 2--3 is lost 
for a 200 GeV neutralino, from either the Sun or the Earth.  It is 
thus highly desirable to keep the threshold as low as possible to keep 
the signal high.

\begin{figure}[!t]
\centerline{\epsfig{file=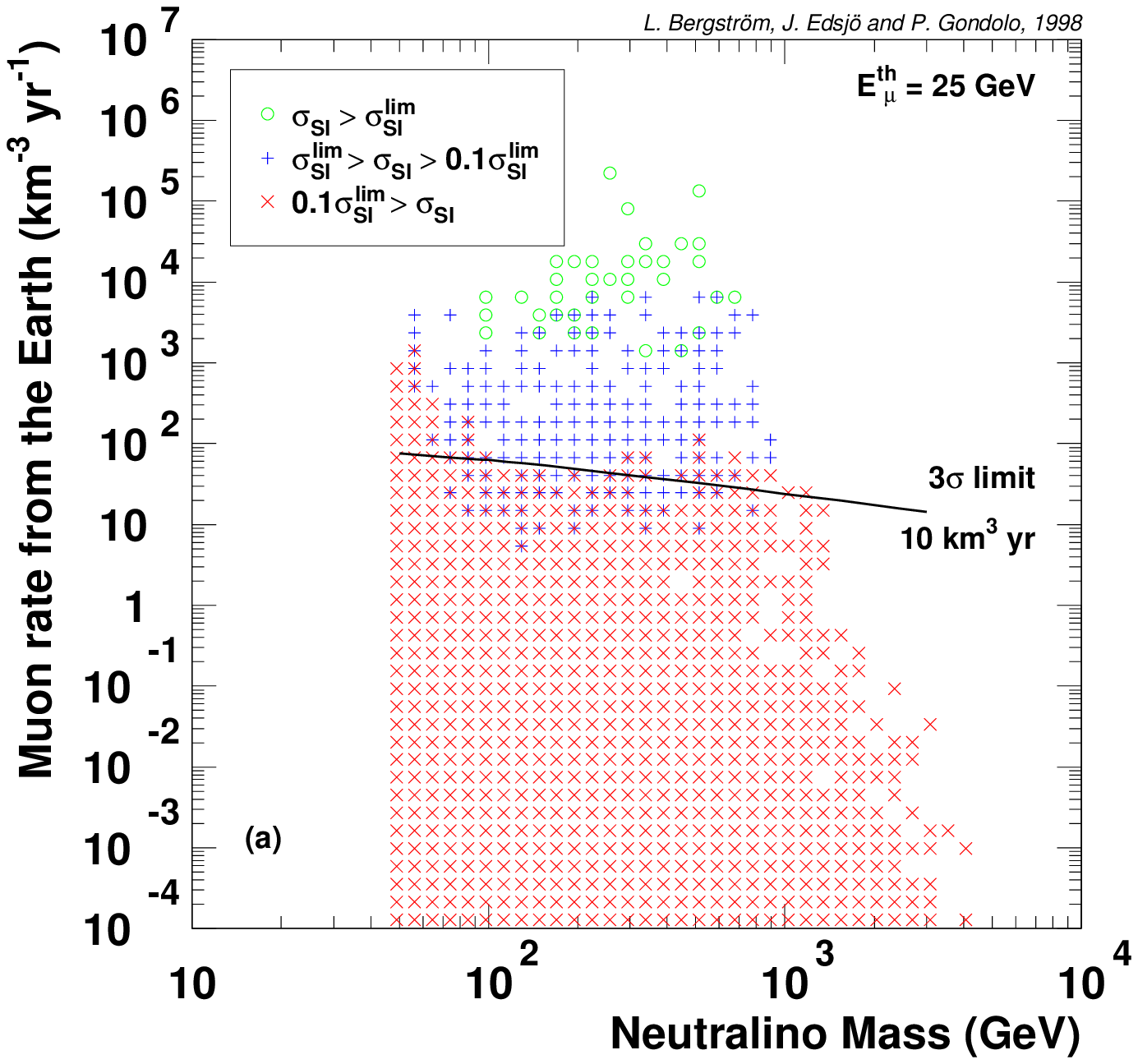,width=0.49\textwidth}
\epsfig{file=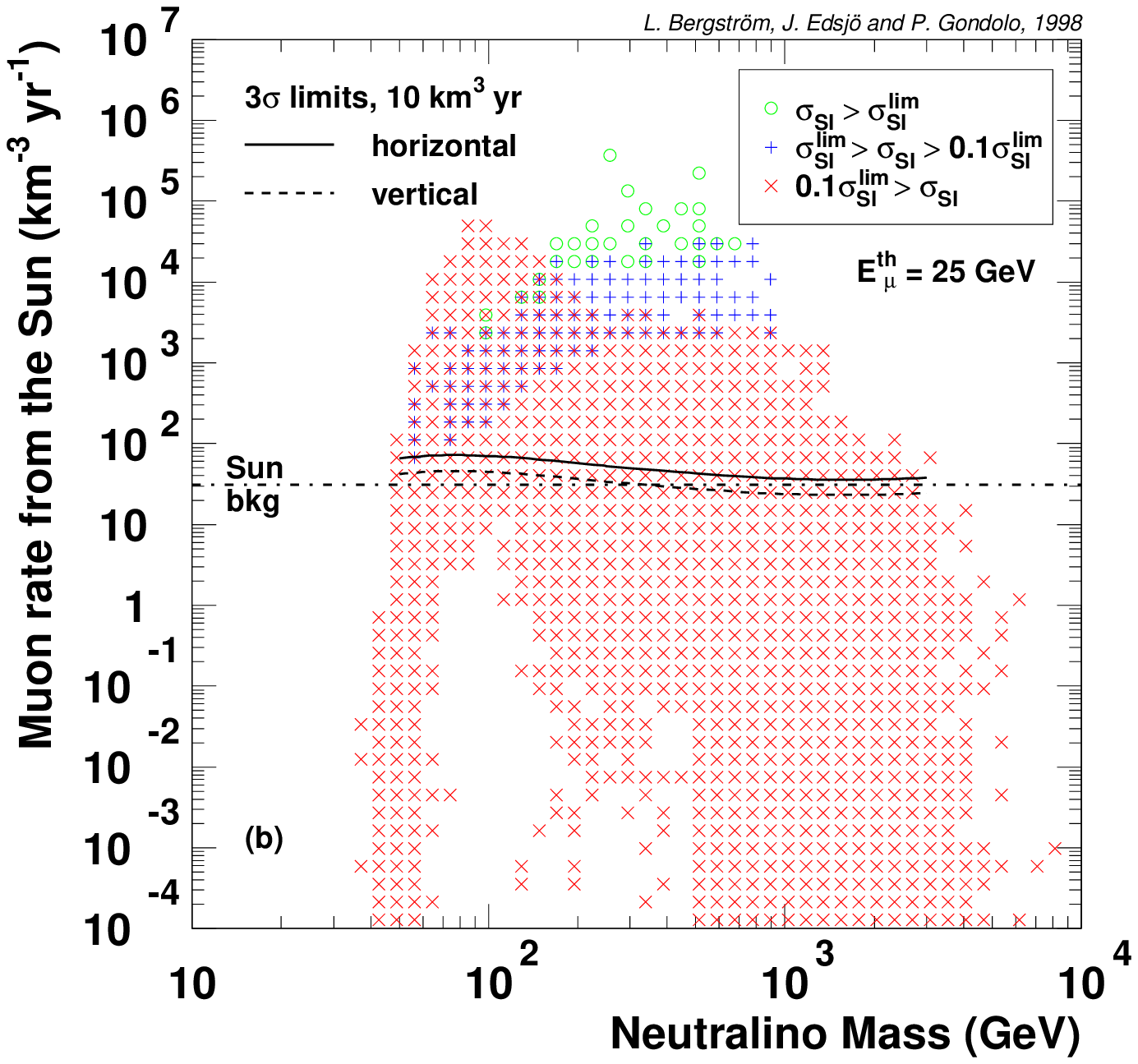,width=0.49\textwidth}}
\caption{The contained event rates in (a) the Earth and (b) the Sun
with a muon energy threshold of 25 GeV\@. The best $3\sigma$ limits
with an exposure of 10 km$^3$ yr are also shown as well as the
background from the Sun's corona.}
\label{fig:rceasumx}
\end{figure}

The above discussion is true for fluxes impinging on a thin detector.
However, for $\cal O($1~km$^{3})$ neutrino telescopes, we would expect
the event rates for contained events (i.e.\ tracks starting inside the
detector) to be high also for masses below a few hundred GeV\@. In
Fig.~\ref{fig:rceasumx} we show the contained event rates for a muon
energy threshold of 25 GeV\@. We clearly see that we can gain at least
an order of magnitude at low masses.  We should however keep in mind
that the background rate also increases and in the figure we show the
best $3\sigma$ limits that can be obtained with an exposure of 10
km$^3$ yr. Comparing with Figs~\ref{fig:reamx} (b) and \ref{fig:rsumx}
(b), we see that below about 300 GeV, the contained events are
expected to be better than the impinging fluxes.  With fully
contained events (i.e.\ tracks both starting and ending inside the
detector) it would be possible to get some information on the muon
energy from the track length which would make it possible to gain up
to a factor of two in sensitivity.  A detailed study of this requires
a detailed detector simulation.

\section{Comparison with other signals} \label{sec:othersignals}

We discuss here if the complementarity between high energy neutrino
searches from the Sun/Earth and other neutralino dark matter searches
 persists at high energy thresholds.

\subsection{Direct detection}

The uncertainties in the capture rates governing the muon flux enter
in a similar way in the calculation of the rates of direct detection
(in particular, the local halo density plays the same role).
Therefore, it is meaningful to make a comparison between indirect and
direct detection \cite{sadoulet}.  To illustrate this point, we show
in Figs.~\ref{fig:reamx}--\ref{fig:rsumx} with different symbols (and
colours) which models are presently ruled out by the most sensitive
direct detection experiments, the {\sc Dama} dark matter search
\cite{dama} and the Gotthard Ge experiment \cite{gotthard}.  We have
also shown which models it should be possible to probe with a factor
of 10 increase in sensitivity of the direct detection experiments.
Most of the models excluded by Baksan are also excluded by the direct
detection experiments under the same set of assumptions.  The
explanation of this is that the spin-independent cross section of halo
neutralinos which cause capture in the Earth is also responsible for
the signal in the {\sc Dama} detector.

\begin{figure}[!t]
\centerline{\epsfig{file=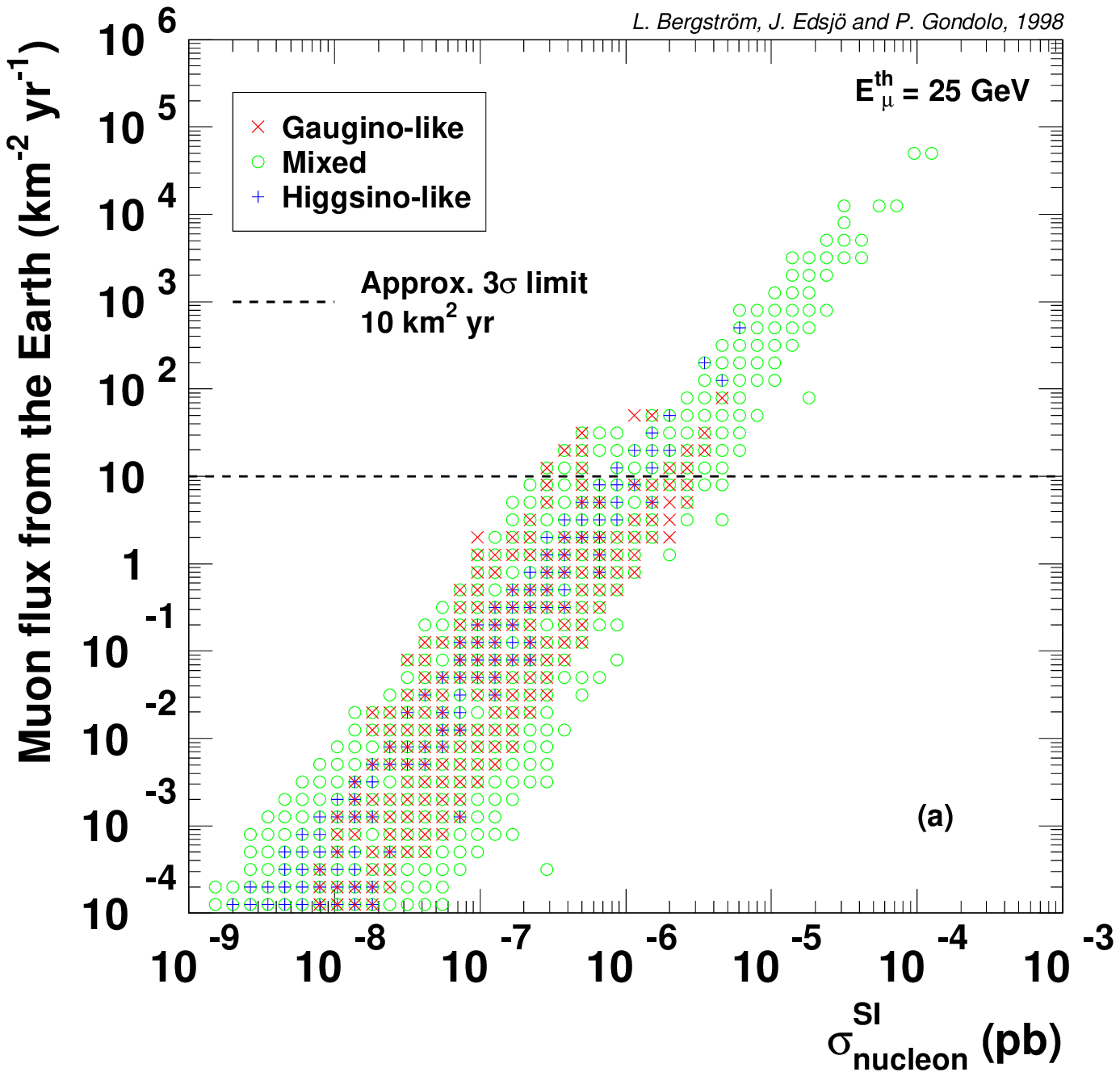,width=0.49\textwidth} 
\epsfig{file=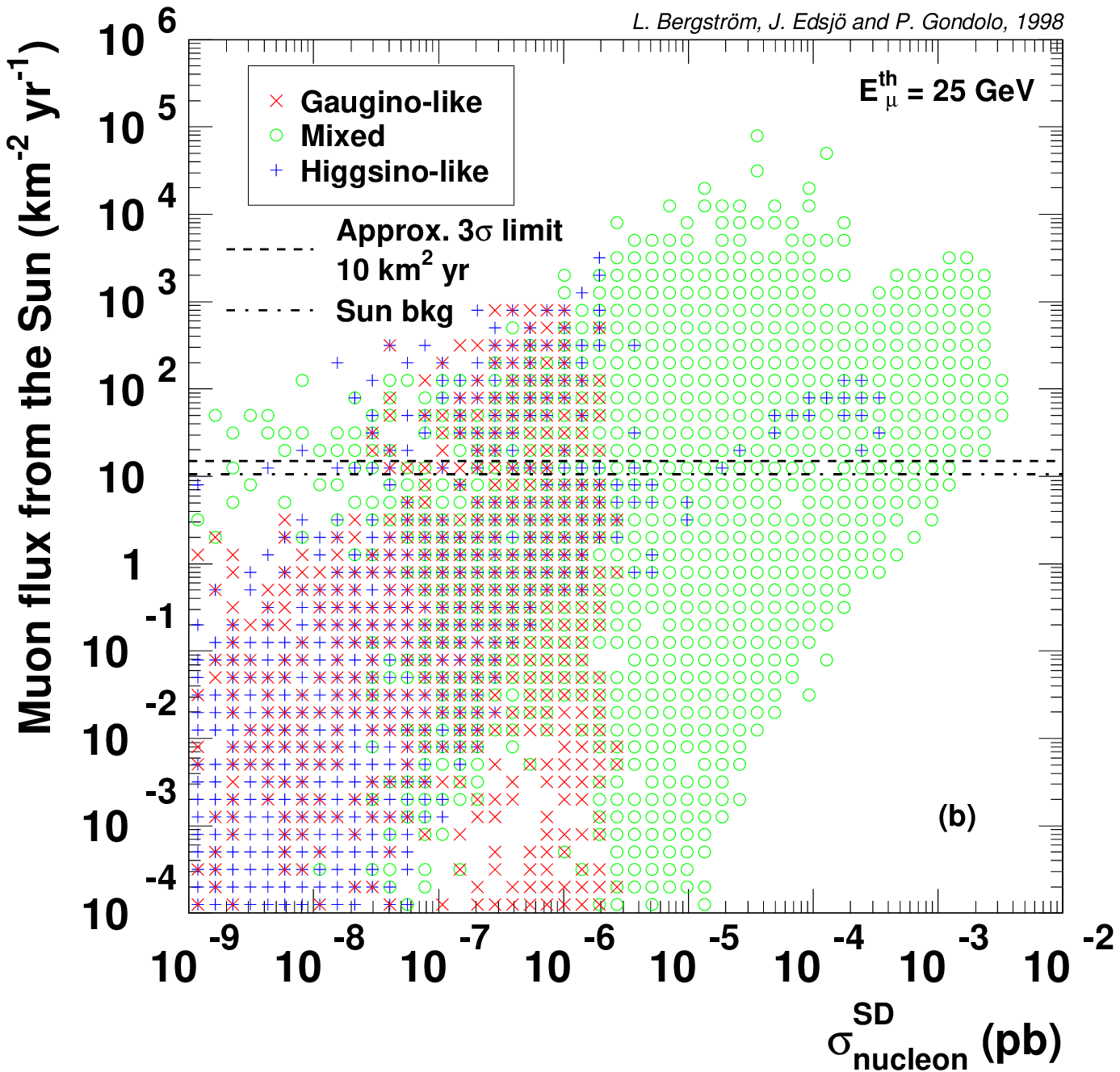,width=0.49\textwidth}}
\caption{The muon fluxes in (a) the Earth versus the spin-independent 
$\chi$-nucleon scattering cross section and (b) the Sun versus the 
spin-dependent $\chi$-nucleon scattering cross section. The 
approximate $3\sigma$ limits that can be obtained with a 10 km$^{2}$ 
yr exposure are also shown.}
\label{fig:reasusisd}
\end{figure}

For the case of muons from the Sun, the situation is quite different, 
however.  Since the Solar medium is dominated by protons, which have 
spin, it is the spin-dependent cross section which is crucial.  As the 
present experimental limits on spin-dependent cross sections are 
several orders of magnitude less restrictive than the spin-independent 
ones, neutrino telescopes provide more stringent bounds at the moment 
through the absence of a neutrino-induced muon signal from the 
direction of the Sun.

Even if one would manage to build a large spin-dependent detector, the
correlation between the signal in such a detector and a muon signal
from the Sun would not be as strong as in the case of spin-independent
detection versus muons from the centre of the Earth.  This is apparent
by comparing Figs.~\ref{fig:reasusisd} (a) and~\ref{fig:reasusisd}
(b).  Fig.~\ref{fig:reasusisd} (a) shows the muon flux from the Earth
versus the spin-independent $\chi$-nucleon cross section,
Fig.~\ref{fig:reasusisd} (b) shows the muon flux from the Sun versus
the spin-dependent $\chi$-nucleon cross section. A strong correlation
is present in the first, a very weak correlation in the second.  (The
correlation between the muon rate from the Sun and the
spin-independent cross section, which we do not show, is even weaker.)
We also show the approximate (neglecting the neutralino mass
dependence) $3\sigma$ limits that can be obtained with an exposure of
10 km$^2$ yr.

In these figures we also show the composition of the respective
neutralinos, where we define a neutralino as being gaugino-like if the
summed fraction of photino and zino components is larger than 0.99, as
Higgsino-like if the fraction is less than 0.01, and as mixed if it
lies between these limits.  As can be seen, in both the case of the
Earth and the Sun, mixed neutralinos give the largest rates, both for
muons and direct detection.  It is also obvious that the correlation
between the two types of detection is very strong for the Earth.  For
the Sun, however, there are several models which give large muon rates
but have small spin-dependent cross sections.

The correlation between large muon rates from the Earth and direct 
rates due to spin-independent scattering is even stronger at a 
threshold of 25 GeV (see  Fig.~\ref{fig:reamx} (b)).  Neutrino telescopes 
may still have an edge at the highest masses (TeV range), where the 
atmospheric neutrino background is also smaller.  The situation for 
the Sun is more favourable for neutrino telescopes, as can be seen in 
Fig.~\ref{fig:rsumx} (b).  Heavy models giving more than $10^2$ 
events per km$^2$ per year may have spin-dependent cross sections down
to $10^{-7}$ pb (and spin-independent cross sections even 
smaller).

\subsection{Continuum $\gamma$--rays from annihilation in the halo}

We write the flux of continuum gammas as
\begin{equation}
  \Phi_\gamma (E,\ell,b,\Delta\Omega) 
  \simeq 1.87 \cdot 10^{-8} 
  \, \frac{d{\cal S}_{\gamma}}{dE}(E) \,
  J(\ell,b,\Delta\Omega)
  \;\;\rm{cm}^{-2}\;\rm{s}^{-1}\;\rm{sr}^{-1}.
  \label{eq:flux}
\end{equation}
$J(\ell,b,\Delta\Omega)$ contains the integration of the $\gamma$
emission along the line of sight, and depends on the halo model.
$\ell$ and $b$ are the galactic longitude and latitude respectively,
and $\Delta\Omega$ is the solid angle over which we integrate the
flux.  For an isothermal sphere with a core radius of 3.5 kpc, our
galactocentric distance of 8.5 kpc and $\Delta \Omega = 1 {\rm~deg}^2, $
we get $J(\ell,b,\Delta\Omega) \simeq 30$ in the direction of the
galactic centre and $ \simeq 0.38 $ in the opposite direction.  For
steeper halo profiles, $ J(\ell,b,\Delta\Omega) $ in the direction of
the galactic centre may be orders of magnitude larger.  In the figures
we set $ J(\ell,b,\Delta\Omega) = 1.3$ which is the average over $\pi$~sr
towards $b=90^{\circ}$ for the isothermal sphere.

\begin{figure}[!t]
\centerline{\epsfig{file=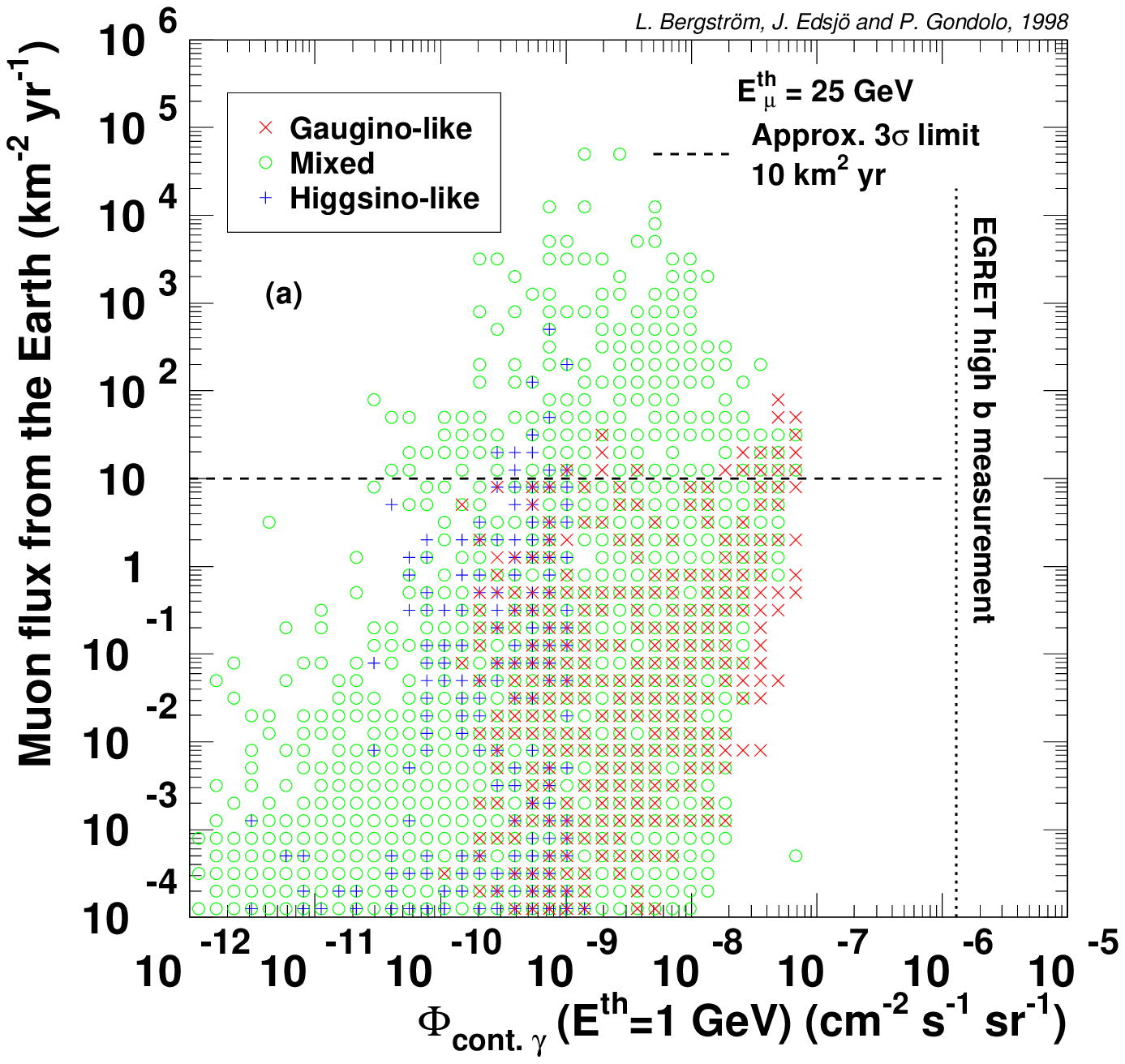,width=0.49\textwidth}
\epsfig{file=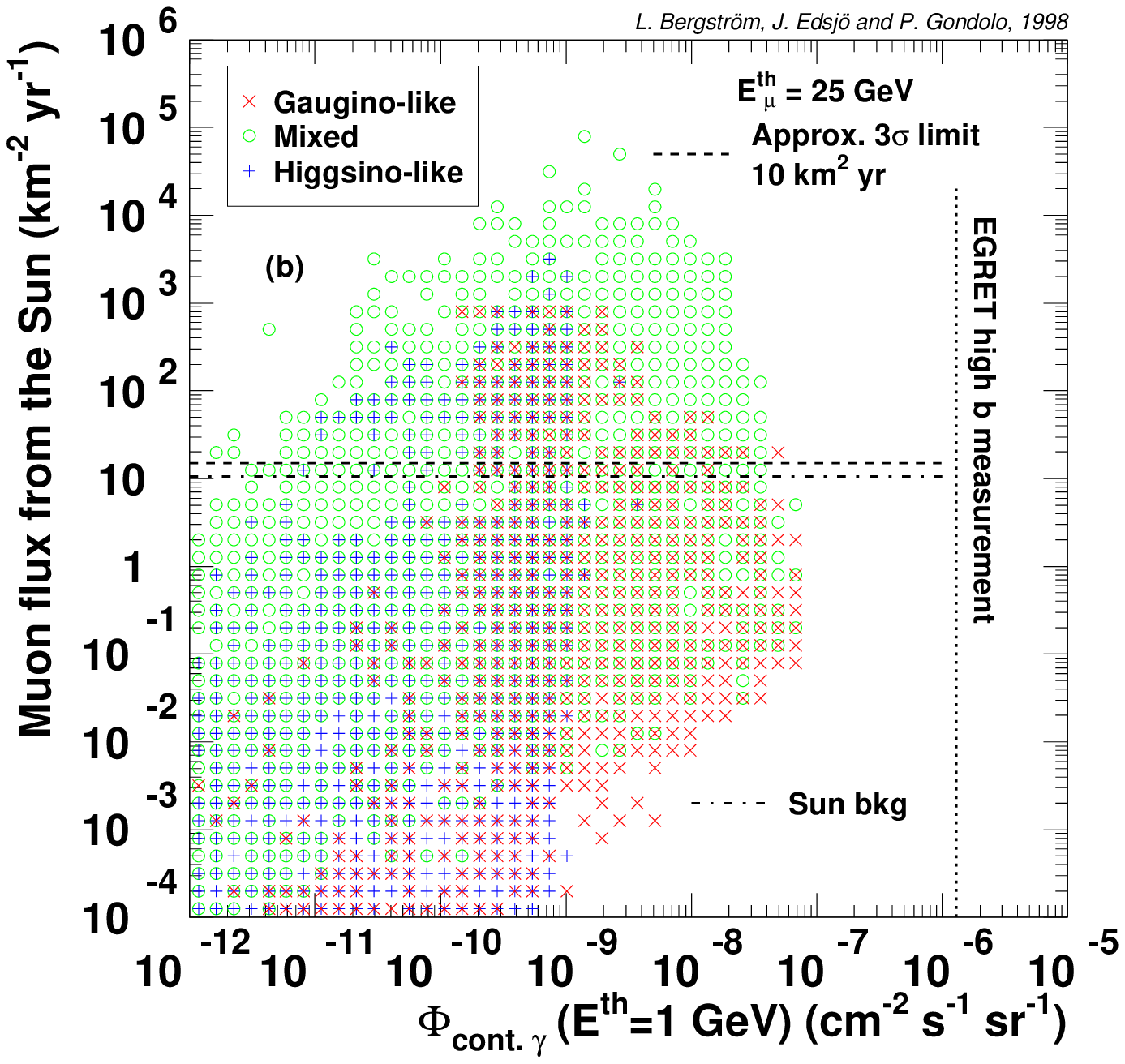,width=0.49\textwidth}}
\caption{The muon fluxes in (a) the Earth and (b) the Sun versus the 
flux of continuum $\gamma$s above 1 GeV from neutralino annihilations 
in the galactic halo.  Also shown is the {\sc Egret} high altitude 
measurement \protect\cite{kniffen} of the gamma ray background.}
\label{fig:reasupgac}
\end{figure}

The other factor in Eq.~(\ref{eq:flux}),
\begin{equation}
\frac{d{\cal S}_{\gamma}}{dE}(E) \simeq  \left( \frac{10\,\rm{GeV}}{m_\chi}\right)^2
  \, \left( \frac{\sigma v}{10^{-26}\ {\rm cm}^3 
   {\rm s}^{-1}}\right)\,\frac{dN_\gamma}{dE} \, , 
\end{equation}
contains the dependence on neutralino properties.  Here $\sigma v$ is 
the annihilation rate and $dN_{\gamma}/dE$ is the energy distribution 
of continuum gammas per annihilation.  The hadronization and/or decay 
of the annihilation products are simulated with {\sc Pythia} 6.115 
\cite{pythia} letting pions and kaons decay.

To illustrate the complementarity of methods, we show in 
Fig.~\ref{fig:reasupgac} (a) the predicted muon flux from the Earth 
versus the flux of continuum gamma rays above 1 GeV, $\Phi_{{\rm 
cont.}~\gamma}$, from annihilations in the halo and in 
(b) the corresponding results for the Sun.  As can be seen, these two 
methods are truly complementary, in particular for the case of the 
Earth (and therefore also when comparing continuum gammas to direct 
detection).  One should note, however, that the predictions for actual 
$\gamma$ ray detectors are much more uncertain than the muon rates due 
to uncertainties in the average halo density profile versus viewing 
angle \cite{bub} and the possibility of clumps of dark matter in the 
halo \cite{beu,begu}.  Also shown in Fig.~\ref{fig:reasupgac} is the 
high-latitude {\sc Egret} limit \cite{kniffen}.

\begin{figure}[!t]
\centerline{\epsfig{file=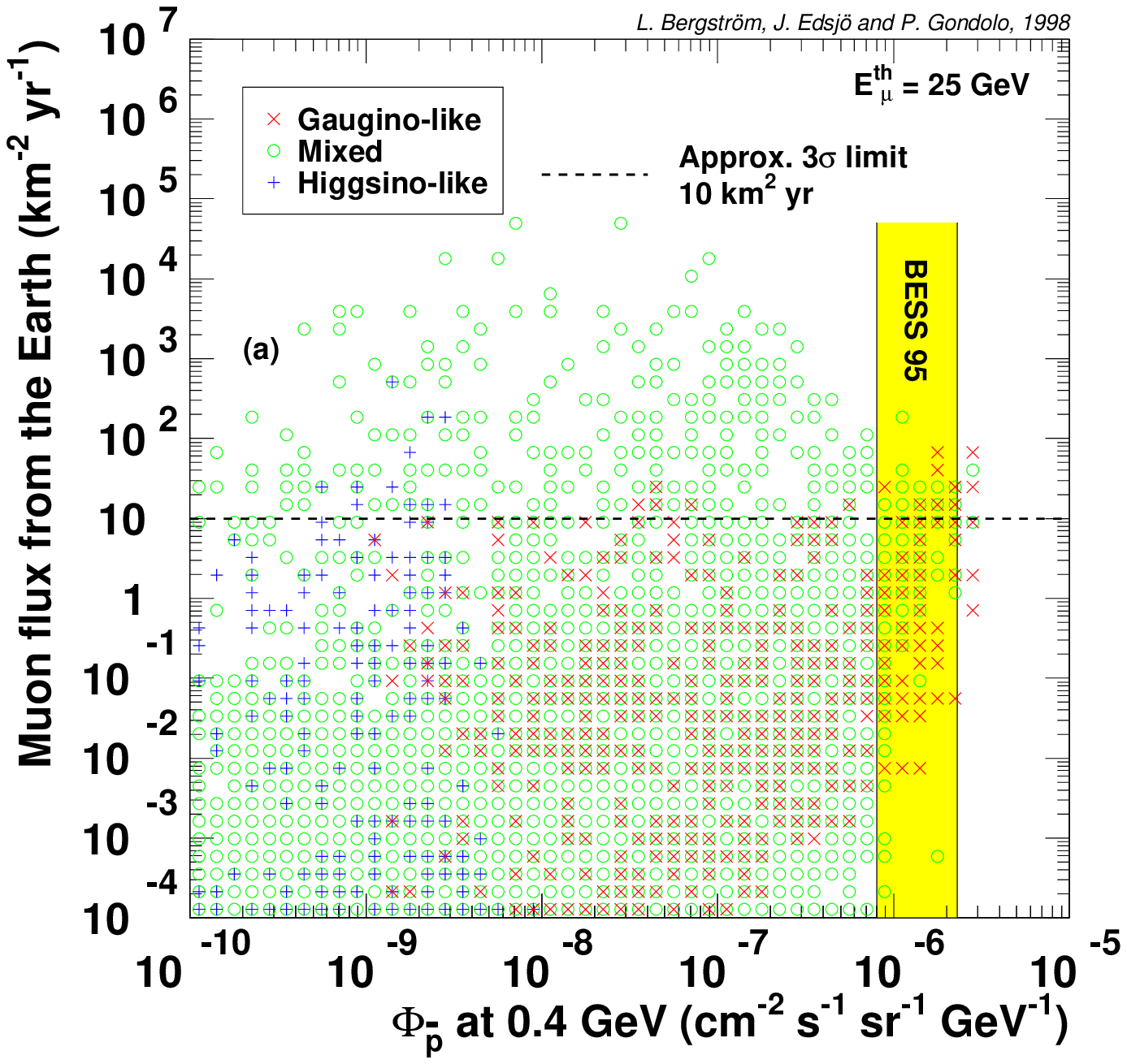,width=0.49\textwidth}
\epsfig{file=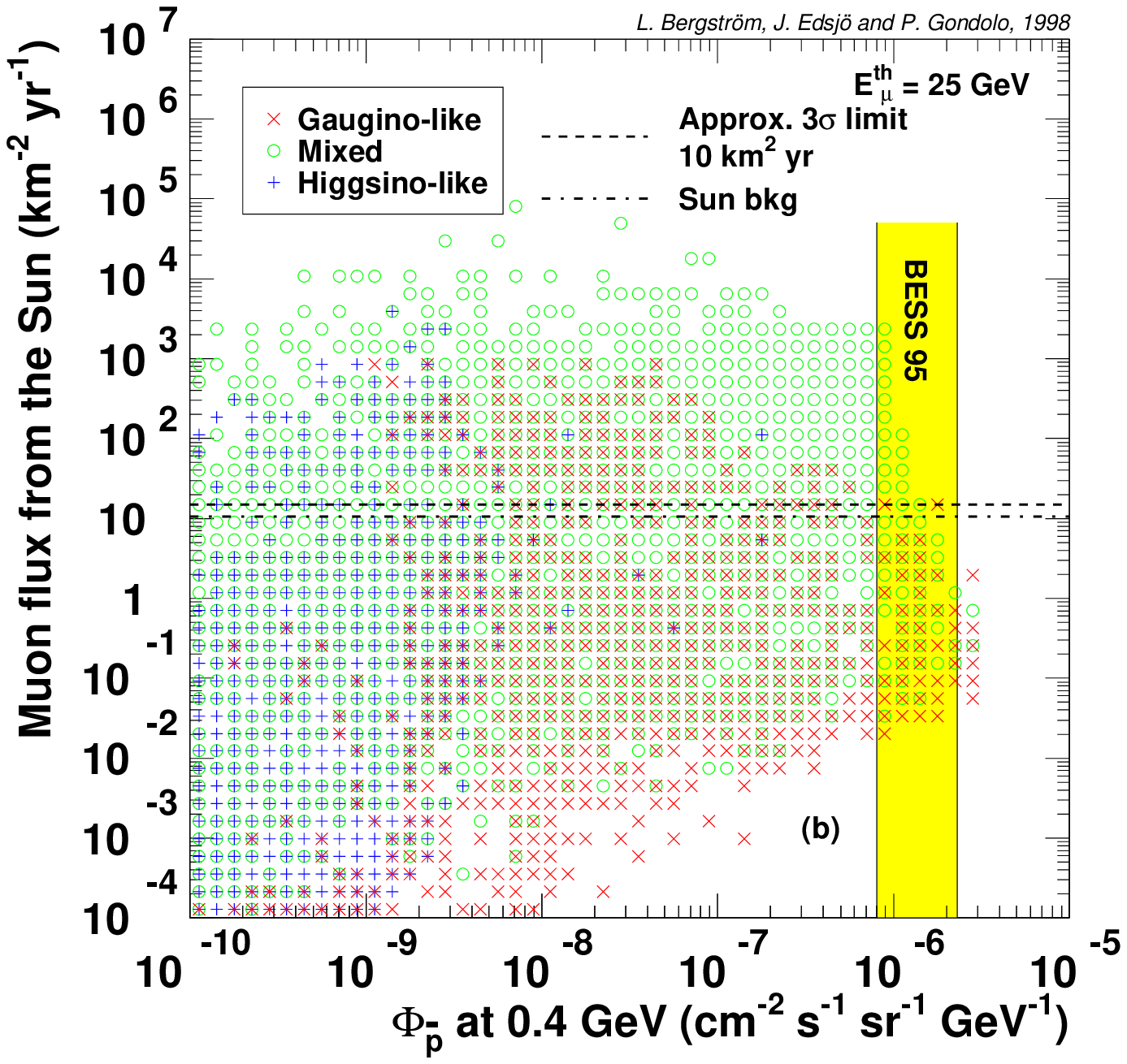,width=0.49\textwidth}}
\caption{The muon fluxes in (a) the Earth and (b) the Sun versus
the flux of antiprotons from neutralino annihilations in
the galactic halo. The {\sc Bess} measurement at 400 MeV 
\protect\cite{bess} is shown as the yellow band.}
\label{fig:reasupbar}
\end{figure}

\subsection{Antiprotons from annihilation in the halo}

For antiprotons, the procedure is very similar, but we now also 
let antineutrons decay in the {\sc Pythia} simulations.  We then use 
the leaky box propagation model with the energy dependent escape time 
given in Ref.\ \cite{salati} and the solar modulation model of 
Ref.\ \cite{perko}.

Astrophysical uncertainties also pertain to antiprotons from 
annihilations in the halo.  Using standard estimates \cite{salati,beu} 
of the antiproton flux at 400 MeV kinetic energy, where the new BESS 
measurements exist \cite{bess}, we find the results shown in 
Fig.~\ref{fig:reasupbar} (a) and (b) where the antiproton flux at 400 
MeV is compared to muons from the Earth and the Sun, respectively.
As can be seen, the BESS experiment is probing some models which 
cannot easily be probed in neutrino telescopes in the foreseeable 
future, neither through muons from the Earth nor from the Sun.  One 
should again be reminded, however, about the large uncertainties in 
the antiproton flux (related, e.g., to the propagation properties in 
the disk and the halo).  Also, since antiprotons do not give as 
clear-cut signature  as muons from the Earth and the Sun
 (for instance, there is no directional 
information), and since the 
experiments are already close to the background level predicted by 
cosmic ray production of antiprotons, it is not clear by how large a 
factor the antiproton limits on neutralinos can be improved.

\subsection{Positrons from annihilation in the halo}

We have also calculated the flux of positrons from neutralino
annihilation in the halo. The simulation procedure with {\sc Pythia}
is very similar to the continuum $\gamma$ and antiproton cases. The
propagation is more difficult though since positrons lose energy. We
have used the propagation model in Ref.\ \cite{kamturner} with an
energy dependent escape time (a more detailed analysis is in
preparation \cite{ebpos}). Comparing with the HEAT measurement at 10
GeV \cite{heat}, we find that our fluxes are an order of magnitude or
more smaller than the measured flux. Given the large uncertainties in
the propagation, it is certainly possible to obtain fluxes that would
give rise to an observable effect.

\subsection{Accelerator searches}

\begin{figure}[!t]
\centerline{\epsfig{file=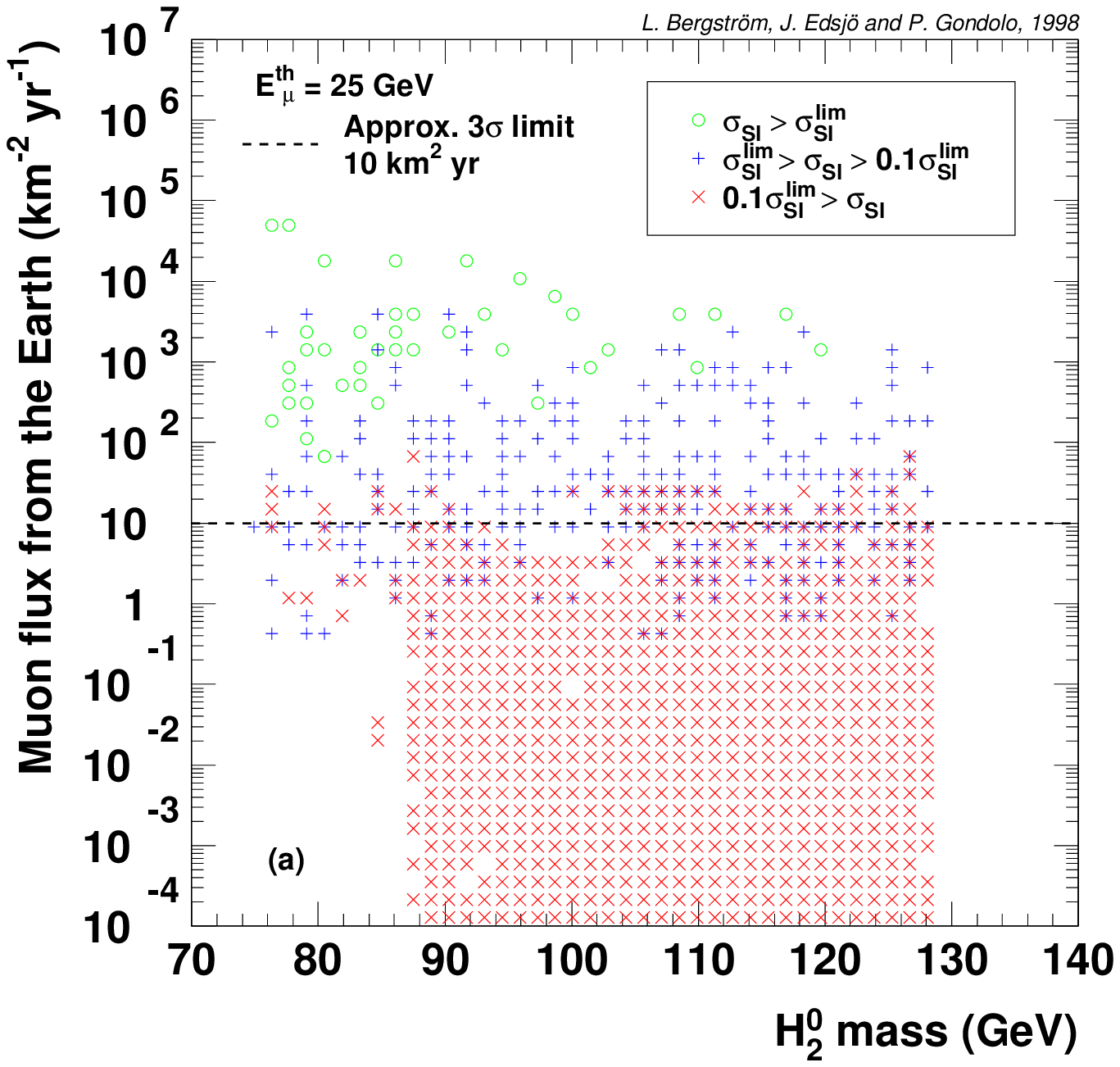,width=0.49\textwidth}
\epsfig{file=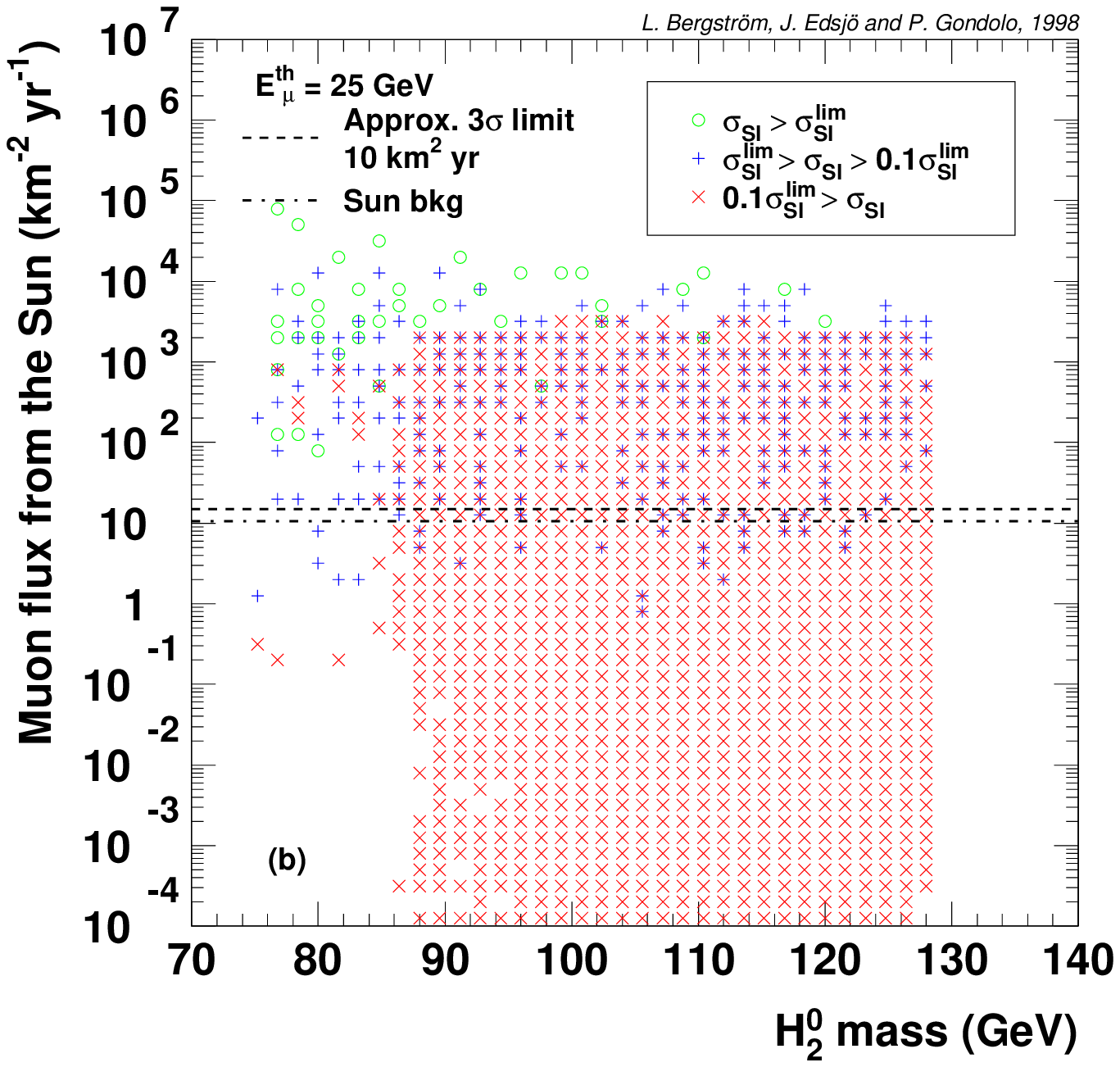,width=0.49\textwidth}}
\caption{The muon fluxes from (a) the Earth and (b) the Sun versus the
lightest Higgs boson mass, $m_{H_2^0}$.}
\label{fig:reasumh2}
\label{fig:lastfig}
\end{figure}

We finally discuss complementarity with respect to accelerator 
searches.  One of the earliest precursors of the MSSM may be the 
discovery of the Higgs boson, where the lightest neutral Higgs scalar 
$H_{2}^0$ in supersymmetric models is limited from above by the $Z^0$ 
mass at tree level, and hardly can be heavier than 130 GeV after loop 
corrections \cite{carena} have been included.  We find models with 
high muon rates all the way up to the heaviest $H_{2}^0$ allowed in 
the MSSM, as exemplified in Fig.~\ref{fig:reasumh2}.  The situation is 
very similar for other thresholds.  An MSSM Higgs boson of mass near 
the 130 GeV limit will not be detectable by LEP II, and may require 
several years of LHC running for its discovery.

\section{Conclusions}

In this paper we have presented extensive calculations of the indirect 
detection rates of neutrino-induced muons from the centre of the Earth 
and the Sun, originating from the annihilation of gravitationally 
trapped neutralinos.  In particular, we have investigated the role of 
the higher muon energy threshold that the next generation of 
kilometer-scale neutrino telescopes is likely to have.  As expected, 
the increased threshold gives reduced rates for the low-mass 
neutralinos whereas the suppression is less severe for high-mass 
models.  For muons from the Earth, the suppression means that neutrino 
telescopes will have some difficulties to compete with direct 
detection methods.  For the Sun the situation is different as the 
spin-dependent cross section has a larger spread, and there do not yet 
exist direct detectors of large sensitivity.  From the point of view 
of neutralino search, the optimum design of a neutrino telescope would 
yield a low muon energy threshold and a good sensitivity to search for 
a signal from the direction of the Sun.

We have shown that the various methods of detecting supersymmetric 
dark matter probe complementary regions of parameter space, and are 
therefore all worth pursuing experimentally.  The dark matter problem 
remains one of the outstanding problems of basic science.  Maybe the 
first clues to its solution will come from the large new neutrino 
telescopes presently being planned.

\acknowledgments

LB was supported by the Swedish Natural Science Research 
Council (NFR).  We thank Piero Ullio for discussions.  This work was 
supported with computing resources by the Swedish Council for High 
Performance Computing (HPDR) and Parallelldatorcentrum (PDC), Royal 
Institute of Technology.


\end{document}